\documentstyle[pre,aps,floats]{revtex}
\input epsf
\begin{document}
\twocolumn[
\title{\bf
Refined Simulations of the Reaction Front for Diffusion-Limited
Two-Species Annihilation in One Dimension.}
\author{Stephen J.\ Cornell\cite{curad}}
\address{
D\'epartement de Physique Th\'eorique, Universit\'e de Gen\`eve\\
24 quai Ernest-Ansermet, CH-1211 Gen\`eve 4, Switzerland.}
\date{UGVA/DPT 1994/10--857: October 15, 1994}
% copied from aps.sty:
\widetext\leftskip=0.10753\textwidth \rightskip\leftskip
\begin{abstract}
Extensive simulations are performed of the diffusion-limited reaction
A$+$B$\to 0$ in one dimension, with initially separated reagents.  The
reaction rate profile, and the probability distributions of the separation
and midpoint of the nearest-neighbour pair of A and B particles, are all
shown to exhibit dynamic scaling, independently of the presence of
fluctuations in the initial state and of an exclusion principle in the model.
The data is consistent with all lengthscales behaving as $t^{1/4}$ as
$t\to\infty$.  Evidence of multiscaling, found by other authors, is
discussed in the light of these findings.
\end{abstract}
\draft
\maketitle
]
\narrowtext
\section{Introduction}
There has been a lot of recent interest in the scaling behaviour of the
reaction front that exists between regions of
initially separated reagents A and B that
perform Brownian motion and annihilate upon contact according to the reaction
scheme A$+$B$\to 0$\cite{gara,%
koliko,jieb,chdr,codrch91,koko,tahakitrwe,arhalast,bnre,laarhast1,laarhast2,%
codrch92,takohakowe,chdrkara,codr93,alhs,blee,co94}.
The evolution of the particle densities $a(x,t)$ and
$b(x,t)$ at position $x$ and time $t$
is governed by the equations
\begin{eqnarray}
{\partial a\over\partial t}&=&D{\partial^2 a\over\partial^2 x} - R,\nonumber\\
\label{eom}
\\
{\partial b\over\partial t}&=&D{\partial^2 b\over\partial^2 x} - R,\nonumber
\end{eqnarray}
where $R$ is the reaction rate per unit volume, and the diffusion
constant $D$ has been assumed equal for both species.
For the Boltzmann equation ansatz $R=kab$, the solution to the
resulting partial differential equations with the initial condition
\begin{equation}
a(-x,0)=a_0\theta(x)=b(x,0)\label{initc},
\end{equation}
where $\theta$ is the Heavyside function, has the scaling property
\begin{equation}
R=t^{-\alpha-(1/2)}\Phi\left({x\over t^\alpha}\right)\quad \hbox{ for }
x\ll t^{1/2} \label{Rscaling},
\end{equation}
with $\alpha={1\over 6}$\cite{gara}.  This result may
be understood by considering the steady-state solutions to Eqs.\ (\ref{eom})
for boundary conditions $a(-x)\to J |x|/D$, $a(x)\to 0$, $b(-x)\to 0$ and
$b(x)\to Jx/D$ as $x\to\infty$.  Under these conditions, there are
opposing constant currents $J$ of either species, and it can be shown
\cite{bnre,codr93}
that
the resulting reaction profile is of the form
\begin{equation}
R_{ss}=J^{1+\lambda}\Phi_{ss}\left({x J^{\lambda}}\right)\label{RscalingSS},
\end{equation}
with $\lambda={1\over 3}$
Returning to the the time-dependent case, the
quantity $(a-b)$ obeys a diffusion equation,
whose solution for initial conditions (\ref{initc}) is
\begin{equation}
a-b = {2a_0\over\sqrt{\pi}}\int_0^{x/(2\sqrt{Dt})} \exp\left(-{y^2}\right)dy.
\label{erf}
\end{equation} Let us assume that the reaction takes place
within a region of width $w\sim t^\alpha$, with
$\alpha<{1\over 2}$.
The profiles for $w\ll x\ll t^{1/2}$ are of the form
$a\propto a_0x/t^{1/2}$, so there is a current of particles arriving the
origin of the form $J=D\partial_x a\sim t^{-{1/2}}$.  The characteristic
timescale on which this current varies is $(d\log J/dt)^{-1}\propto t$, whereas
the equilibration time for the front is of order $ (w^2/D)\sim t^{2\alpha}
\ll t$.  The front is
therefore formed quasistatically, and so Eq.\ (\ref{Rscaling}) may be
obtained from (\ref{RscalingSS}) simply by writing $J\propto t^{-{1/ 2}}$.

Simulations and experiments appear to confirm these results when the spatial
dimension $d$ is two or greater\cite{koliko,jieb,chdr,koko}.
In dimension less than two, strong
correlations between the motions of the two species
cause the Boltzmann approximation $R=kab$ to break down.
However, the solution
to the steady state problem is still of the form (\ref{RscalingSS}), albeit
with a different exponent $\lambda = 1/ (d+1)$\cite{codr93,blee}.
If the results from the
steady-state may still be used, this would lead again to dynamical scaling
of the form (\ref{Rscaling}), with $\alpha={1\over 4}$ in $d=1$.
Simulations using a one-dimensional Probabilistic Cellular Automata (PCA)
model appeared to verify the dynamical scaling form (\ref{Rscaling}), though
with $\alpha=0.293\pm 0.005$\cite{codrch91}.
Monte-Carlo simulations also found
$\alpha\approx 0.30\pm 0.01$\cite{arhalast}.

However, a recent article by Araujo {\it et al\/} \cite{alhs}
has challenged the validity of the scaling form (\ref{Rscaling}).
This article reported Monte Carlo (MC) simulations in one dimension, using
an algorithm where the A and B particles always react on contact and so
are unable to cross over each other.  The right-most A particle (RMA) is
therefore always to the left of the left-most B particle (LMB).  Defining
$l_{AB}$ as the separation between the RMA and LMB, and $m$ as the
midpoint between them, Araujo {\it et al\/} found that
the probability distributions $P_l$ and $P_m$
of these quantities displayed dynamic
scaling, with characteristic lengthscales $t^{1/4}$ and
$t^{3/8}$ respectively.  Meanwhile, the different moments of the
reaction profile were described by a continuous
spectrum of lengthscales between $t^{1/4}$ and $t^{3/8}$.
More specifically, defining
\begin{eqnarray}
l^{(q)}&\equiv&\left(\int_0^\infty l_{AB}^q
P_l(l_{AB})dl_{AB}\right)^{1/q}\\
m^{(q)}&\equiv&\left(\int_{-\infty}^\infty |m|^qP_m(m)dm\right)^{1/ q}\\
x^{(q)}&\equiv& \left({\displaystyle\int_{-\infty}^\infty |x|^qR(x,t)dx
\over\displaystyle\int_{-\infty}^\infty R(x,t)dx}
\right)^{1/ q},
\end{eqnarray}
Araujo {\it et al\/} found that $l^{(q)}\sim t^{1/4}$ and
$m^{(q)}\sim t^{3/8}$, but $x^{(q)}\sim t^{\alpha_q}$ with
${1\over 4}\le\alpha_q<{3\over 8}$ increasing monotonically with $q$.
They also proposed the following form for $R$:
\begin{equation}
R(x,t)\approx t^{-{1/4}}
\left({x\over t^{1/4}}\right)^{-2}\exp\left(-{|x|\over
t^{3/ 8}}\right),\label{Rmulti}
\end{equation}
which predicted values of $\alpha_q$ that were in good agreement with
their numerical findings (the prefactor $t^{-{1/4}}$, essential for
consistency, is missing in \cite{alhs}).
The authors of \cite{alhs} argued that Poisson noise in the initial state
causes the reaction
centre to wander anomalously as $m\propto t^{3/8}$,
invalidating the use of the steady state results.

In this paper, I first describe
extensive simulations of this system, using two independent models---the PCA
model used in \cite{codrch91},
and a MC model similar to that of Araujo
{\it et al} in \cite{alhs}.
I find that dynamical scaling appears to hold for $P_l$, $P_m$,
and $R$, independently of the existence of Poisson fluctuations in the
initial state and of the presence of an exclusion principle.  While I
confirm the result $l^{(q)}\sim t^{1/4}$, I find instead that both
$m^{(q)}$ and $x^{(q)}$ appear to scale as $t^\alpha$, with $\alpha\approx
0.28\pm 0.01$, independently of $q$.
The high statistics and wide time domain accessible in the
PCA simulations show that this exponent is decreasing monotonically in time,
consistent with the asymptotic result $\alpha={1\over 4}$ predicted by
the analogy with the steady-state result.  The measured forms of
$P_l$, $P_m$ and $R$ are found to be described by
very simple analytic forms to high accuracy.
I then
discuss the validity of the fluctuation argument used by Araujo {\it et al\/}
to explain the result $m\sim t^{3/8}$.  An exact
calculation of a related quantity suggests that the wandering of the
reaction centre should instead be of the order $\sim t^{1/4}$, which
is not sufficient to make the use of the steady-state analogy invalid.
Some of these results have been discussed in a previous publication\cite{co94}.

\section{Monte-Carlo Model}
\subsection{Description of Model}
The model described in Ref.\ \cite{alhs} consists of independent random walkers
with no exclusion principle.  In the interests of computational efficiency, I
used a model which is identical provided the site occupation number is
not too large, but whose site updates may be effected using a lookup-table
algorithm.  In this way, it was possible to obtain statistics equivalent
to the simulations in \cite{alhs} in the space of a few days.

The model has an `exclusion principle', in that no more than
$2 l_p$  particles of each type are allowed per site.
In the diffusion step, each of these particles moves onto a
neighbouring site, in such a way that no more than $l_p$ particles may move
from a given site in the same direction at once.
This constraint automatically satisfies the `exclusion principle'.
If there are $l_p$ or less
particles on a site, then the direction in which each particle moves is chosen
independently at random.  If there are more than $l_p$ particles, the
same redistribution method is used for the `holes'---i.e.\ the probability
of $j$ particles moving to the right when the occupation number is
$k$ is the same as $(l_p-j)$ particles moving to the right when the occupation
number is $(2l_p-k)$.
The diffusion constant for this model is $1\over 2$.

In these simulations, the value $l_p=13$ was used (this was the largest
value that could be implemented efficiently).  Since the average density
was 1 or less, the frequency of events where the occupation is greater than
$l_p$ is of order $e^{-1}/(l_p+1)!\approx 4\times 10^{-12}$, so these events
are extremely rare (the simulations represent 21000 samples of 4000 sites
over 25000 timesteps, so the expected total number of such events is
less than 10).  The influence of such events on the results is still smaller,
since the probability of a large number of particles spontaneously moving in
the same direction is low (e.g.\ 14 independent walkers move in the same
direction with probability $2^{-13}\approx 10^{-4}$).  Moreover, the
universality class for the scaling properties is not expected to depend on
such events, as the reactions take place in the zone where the density is
low.  These results may therefore justifiably described as equivalent to
those reported in \cite{alhs}.  The {\sc Fortran}
implementation of this
algorithm performed $1.4\times 10^7$ site updates per second on a
HP 9000/715/75 workstation.

\begin{figure}
\epsfxsize=\hsize
\centerline{\epsfbox{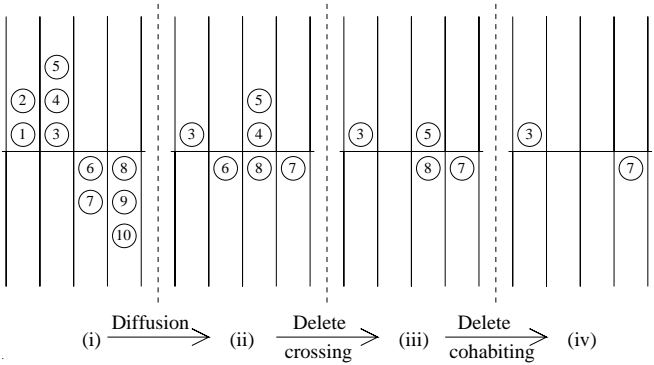}}
\caption{Illustration of the Monte Carlo reaction-diffusion algorithm,
showing the diffusion step and the two stages of the reaction step
that first remove particles that have crossed over and then
react those that are at the same site.\label{mc-alg}}
\end{figure}

One timestep consists of moving all the particles, followed by a
reaction step.
The pure diffusion algorithm has a spurious invariance, in that particles
initially on even sites will always be on even sites after an even number
of timesteps, and will be on odd sites at odd timesteps (and contrarily for
particles initially on odd sites).
In accordance with the prescription in \cite{alhs} that an A particle never be
found to the right of a B particle, it is important that the reaction
takes account of particles of different types crossing over each other
(i.e.\ an A-particle at site $i$ hopping to $i+1$ at the same time that
a B-particle at $i+1$ hops to site $i$).
The reaction algorithm first removed such particles, then
annihilated any remaining A and B particles occupying the same site.
This is illustrated in Fig.\  \ref{mc-alg}.  Four sites are shown,
with initially two sites occupied by A-particles (represented above
the line, labeled 1--5) and two occupied by B-particles (beneath the line,
labeled 6--10).  In the diffusion step, each particles moves onto a
neighbour at random, producing the state (ii).
The reaction first deletes the A-B pair (4,6) that
crossed over, leading to state (iii), then removes the pairs that sit at the
same site, giving a final state (iv).
If the `delete crossing' step were not present, the reaction step would
simply remove the pair (4,8) from state (ii), leading to a state where
there are B-particles to the left of A particles.

\subsection{Simulation Results}

An approximation to a Poissonian initial state of average density unity
was prepared by performing 16 attempts to add an A-particle,
with probability $1/16$,
to each of the first 2000 sites of a 4000-site lattice.
The other half of the lattice was similarly populated with B-particles.
At the boundaries, particles that attempted
to leave the system were allowed to do so, but a random number (distributed
binomially between 0 and 16, average $1\over 2$) of particles
was allowed to re-enter the system at the end sites.
The average density at the extremities was thus kept at the value
unity.

In order to mimic the simulations in \cite{alhs} as closely as possible,
instantaneous
measurements were made of $l_{AB}$, $m$, and the concentration
profiles of the product and  reagents
at times 1000, 2500, 5000, 7500,\dots, 25000.  These were then
averaged over 21000 independent initial conditions.  The quantities $l^{(q)}$
and $m^{(q)}$ were measured from the probability distributions over the
samples, and a quantity $X^{(q)}$ was defined as
\begin{equation}
X^{(q)}=\left({\displaystyle\int x^q C(x,t)\,dx\over
\displaystyle\int C(x,t)\,dx}\right)^{1/q},
\end{equation}
where $C\equiv\int R\,dt$ is the profile of the reaction product.
This quantity differs from $x{(q)}$, but since
$\int C\,dx\propto t^{1/2}$, and provided that $x^{(q)}$ behaves as a
power of $t$, Eq.\ (5) of \cite{alhs} shows that they should have the same
scaling behaviour.

\begin{figure}
\epsfxsize=\hsize
\centerline{\epsfbox{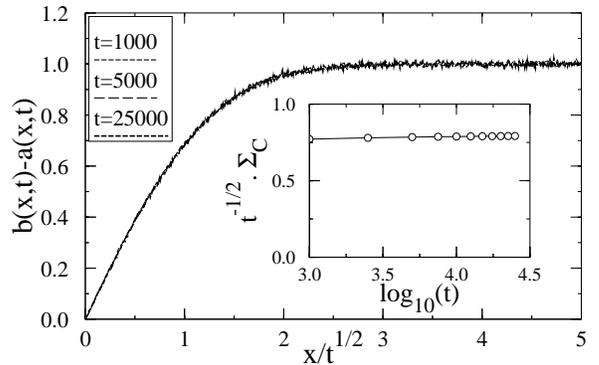}}
\caption{Scaling plot of the density difference $a(x,t)-b(x,t)$ for the
MC data with random initial condition.  Inset: bias plot for the
total number of C particles $\Sigma_C\equiv\int C(x,t)\,dx$.\label{fse-mc}}
\end{figure}

{}From Eq.\ (\ref{eom}), the difference in the particle densities,
$(a-b)$, obeys a simple diffusion equation, whose solution
is given by (\ref{erf}). Any finite-size effects
in the data would first show up in deviations of the particle profiles from
the values they would have for an infinite system.  Figure \ref{fse-mc}
shows a plot of $(a-b)$ as a function of $(x/t^{1/2})$ for three time
values, displaying excellent rescaling.  A simpler test
of finite-size effects
is to show that the total C-particle number $\Sigma_C\equiv\int C\,dx$ is
proportional to $t^{1/2}$.  The inset to Fig.\ \ref{fse-mc}
confirms that  $\Sigma_C\cdot
t^{-{1/2}}$ is indeed independent of time.

\begin{figure}
\epsfxsize=\hsize
\centerline{\epsfbox{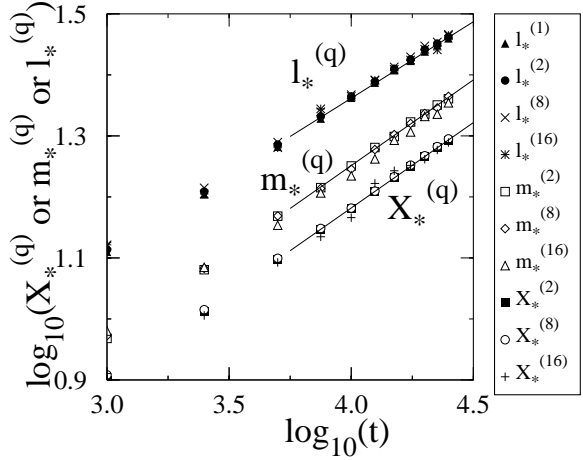}}
\caption{Log-log plot of $l_*^{(q)}$, $m_*^{(q)}$, and $X_*^{(q)}$
(see text) from the MC simulations with Poisson initial conditions.}
\label{moms-mc}
\end{figure}

Figure \ref{moms-mc} is a log-log plot of $X_*^{(q)}$, $m_*^{(q)}$, and
$l_*^{(q)}$
as a function of time, where
\begin{eqnarray}
X_*^{(q)}(t)&\equiv&\xi_q X^{(q)}(t)\label{defXstar}\\
m_*^{(q)}(t)&\equiv&\mu_q m^{(q)}(t)\label{defmstar}\\
l_*^{(q)}(t)&\equiv&\lambda_q l^{(q)}(t)\label{deflstar},
\end{eqnarray}
and $\xi_q$, $\mu_q$, and $\lambda_q$ are constants that will be
defined later.
The straight lines are fits to the last 8 points for $X_*^{(2)}$,
$m_*^{(2)}$, and
$l_*^{(1)}$.  The gradients for least squares fits to
the curves in Fig.\ \ref{moms-mc} are listed in Table \ref{table}.
The exponent describing $l^{(q)}$ is close to $\frac{1}{4}$, as was found in
\cite{alhs}.  However, the results for $m^{(2)}$ and $X^{(2)}$ differ
dramatically from those of Araujo {\it et al}.  Firstly, the exponent
describing $m(t)$ appears to be close to $0.29$, instead of 0.375 as they
found.  Secondly, the exponents describing $X^{(q)}$ appear to be independent
of $q$.  This means that $C(x,t)$, and by implication $R(x,t)$, obeys a simple
scaling form, in contrast to the anomalous form (\ref{Rmulti}).

\begin{figure}
\epsfxsize=\hsize
\centerline{\epsfbox{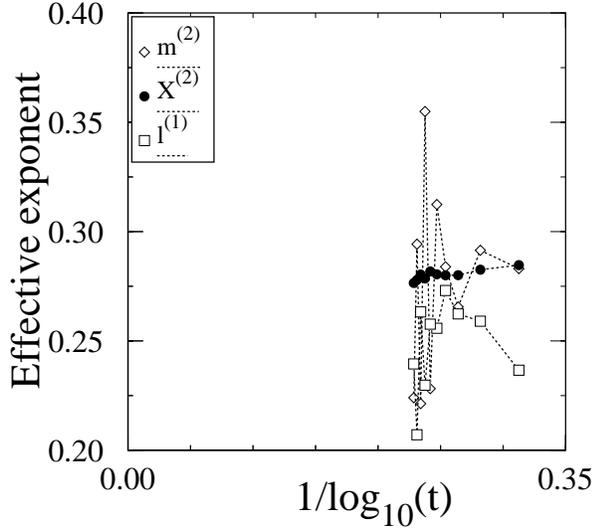}}
\caption{Effective exponents for $l^{(1)}$, $m^{(2)}$, and $X^{(2)}$
(see text) from the MC simulations.}
\label{sg-mc}
\end{figure}

To investigate for a trend in the exponents describing these quantities,
the effective exponent (defined as the gradient between successive points
in Fig.\ \ref{moms-mc}) is plotted as a function of $1/\log_{10}(t)$ in
Fig.\ \ref{sg-mc}.  The data for $l^{(1)}$
and $m^{(2)}$ are far too noisy for any information to be obtained.
The exponent for $X^{(2)}$ appears to
to decrease slowly in time, but the time window in
these simulations is too narrow for conclusive deductions to be made.

\begin{figure}
\epsfxsize=\hsize
\centerline{\epsfbox{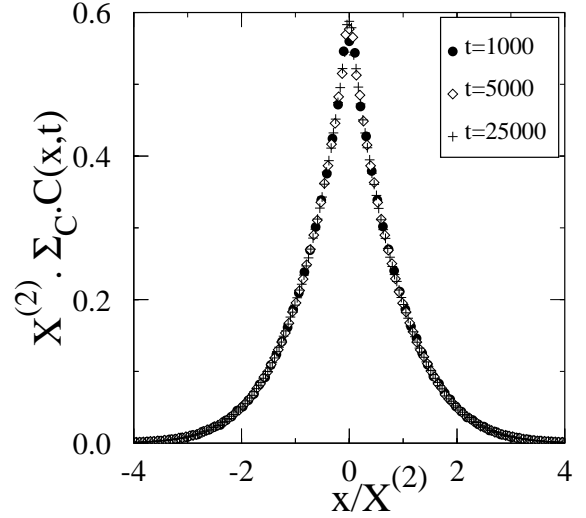}}
\caption{Scaling plot for $C(x,t)$, for the MC simulations.}
\label{proC}
\end{figure}

Figures \ref{proC}, \ref{prol}, and \ref{prom}
are plots of $C$, $P_m(m)$ and $P_l(l_{AB})$, as a function of appropriate
scaling variables, to show the subjective quality of scaling for these
quantities.  The profiles of $P_m(m)$ and $P_l(l_{AB})$ suggest the following
forms:
\begin{eqnarray}
P_m(m)&=& {1\over m_0\sqrt{\pi}}\exp\left[-\left({m\over m_0}\right)^2
\right]\label{pmgauss}
\\
P_l(l)&=& {2l\over l_0^2}\exp\left[-\left({l\over l_0}\right)^2
\right]\label{plgauss}
\end{eqnarray}
These forms predict the following results for the moments of these
distributions:

\begin{figure}
\epsfxsize=\hsize
\centerline{\epsfbox{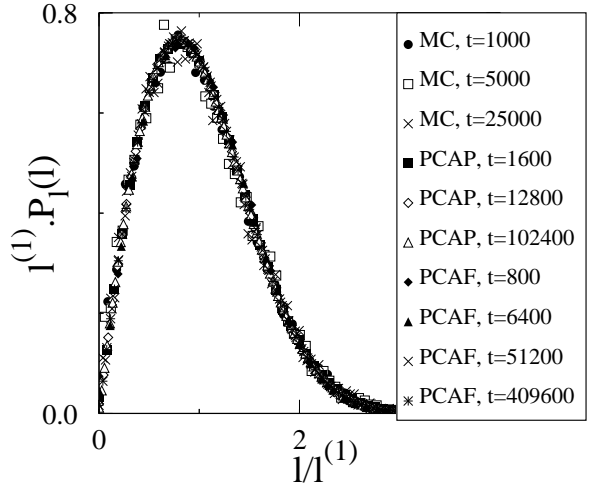}}
\caption{Scaling plot for $P_l(l)$, for the MC simulations (`MC') and the
PCA simulations with Poisson (`PCAP') and full (`PCAF') initial conditions.}
\label{prol}
\end{figure}

\begin{equation}
m^{(q)}=\mu_q^{-1} m_0,\qquad l^{(q)}=\lambda_q^{-1} l_0,
\end{equation}
where
\begin{eqnarray}
\mu_q&=&\left({(q/2)!\over q!}\right)^{1/q}\left[1+(-1)^q\right],
\label{muq}\\
\lambda_q&=&\cases{\mu_{q+1}(\sqrt{\pi}/2)^{1/q}& for $q$ odd,\cr
\left[(q/2)!\right]^{1/q}& for $q$ even.}
\end{eqnarray}
Using these values of $\mu_q$ and $\lambda_q$ in Eqs.\
(\ref{defXstar}--\ref{deflstar}), one would expect $m_*^{(q)}$ and
$l_*^{(q)}$ to be independent of $q$ if the forms
(\ref{pmgauss},\ref{plgauss}) are valid.  The coincidence of the
curves in Fig.\ \ref{moms-mc} confirms this.

Figure \ref{gauss} is an explicit test of the forms (\ref{pmgauss},%
\ref{plgauss}) against the data, by plotting $\log[m^{(2)}P_m(m)]$
and $\log[l^{(2)}l^{-1}P_l(l)]$ against $(m/m^{(2)})^2$ and
$(l/l^{(2)})^2$ respectively, at $t=25000$.  The Y-ordinate has been
shifted so that all curves are coincident at the origin.
The curve labeled $R_{\rm MC}$ is  $\log[C(x,25000)-C(x,22500)]$, which
is approximately proportional to $R(x,25000)$, as a
function of $(x/X^{(2)})^2$.  The straight line for
this curve suggests that the reaction profile $R(x,t)$ is also a Gaussian.
This again contradicts the form (\ref{Rmulti}) proposed by Araujo
{\it et al}.
It is not possible, however, to derive analytical forms for $\xi_q$ that lead
to $X_*^{(q)}$ being independent of $q$ without assuming a form for $x^{(q)}
(t)$ for all $t$,
so the values of $\xi_q$
used in Eq.\ (\ref{defXstar}) were chosen numerically
in an {\it ad hoc\/} fashion.

\section{Probabilistic Cellular Automata Model}
\subsection{Description of Model}
This model has been described extensively in previous
publications\cite{codrch91,chdr91}.
In the one-dimensional realization of this model,
there are up to two particles of  each species at each site, labeled by the
direction from which they moved onto the site at the previous timestep.
The diffusion step consists of changing the velocities of these particles,
then moving the particles onto the neighbouring sites according to their
new velocities.  If there are two particles per site,
they both move in opposite directions, whereas a single particle will
change direction with probability $p$.  The value used in these simulations was
$p=\frac{1}{2}$, so that the particle forgets its
previous velocity at each timestep, and the model is equivalent to the
MC model with $l_p=1$.

\begin{figure}
\epsfxsize=\hsize
\centerline{\epsfbox{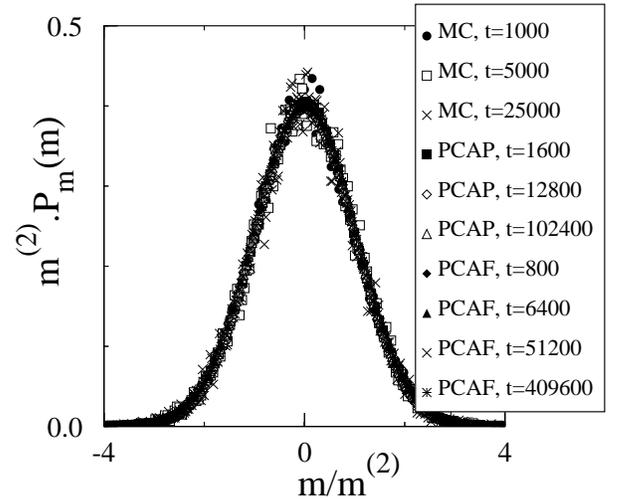}}
\caption{Scaling plot for $P_m(m)$, for the MC simulations (`MC') and the
PCA simulations with Poisson (`PCAP') and full (`PCAF') initial conditions.}
\label{prom}
\end{figure}

The reaction step consists of checking each site for simultaneous occupancy
of A and B particles at the start of the timestep , and removing any
pairs that
hopped onto the site from opposite directions.  Using the segregated initial
condition, and this `infinite' reaction rate, a site can only be occupied by
an A-B pair if the A arrived from the left and the B from the right.
This model has the same two-sublattice structure as the Monte-Carlo
model defined above, and this is preserved by the reaction algorithm,
so these two sublattices must be viewed as two independent systems.
There is therefore an independent nearest-neighbour A-B pair for
each sublattice.  A multi-spin-coding implementation of the
algorithm simulates 64 independent systems at once.

The quantities $P_m(m)$, $P_l(l)$ and $R(x,t)$ at measurement time $t$ were
estimated by averaging over the interval $t(1-\delta)<t<t(1+\delta)$, with
$\delta=0.05$.
We may estimate the order of magnitude of the systematic error that this
introduces into the measured shape of these quantities.
Let $\tilde F(x,t)$ be the
estimate of a function $F(x,t)$ using
the above method.  Then
\begin{eqnarray}\label{syserr}
\tilde F&\equiv& \frac{1}{2t\delta }
\int_{t(1-\delta)}^{t(1+\delta)}F(x,t')\,dt'\\
&=& \frac{1}{2t\delta }
\int_{t(1-\delta)}^{t(1+\delta)}\left(F(x,t)+(t'-t)\frac{\partial}%
{\partial t}F(x,t)\right. \nonumber\\
&&\phantom{\frac{1}{2\delta t}
\int_{t(1-\delta)}^{t(1+\delta)}\Bigl(}\left.
+ {1\over 2}(t'-t)^2\frac{\partial^2}{\partial t^2}F(x,t)
+\dots\right)dt'\\
&=& F(x,t)+ \frac{(t\delta)^2}{6}\frac{\partial^2}{\partial t^2}F(x,t)+{\cal O}
\left[(t\delta)^4\right].
\end{eqnarray}
The fractional error is therefore of order $(t\delta)^2\ddot{F}/(6F)$.

\begin{figure}
\epsfxsize=\hsize
\centerline{\epsfbox{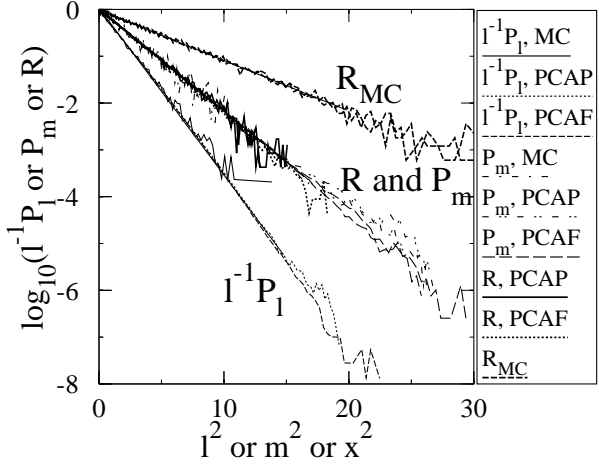}}
\caption{Fits of $P_m$, $P_l$, and $R$ to Eqs.\ (\protect{\ref{pmgauss}}),
(\protect{\ref{plgauss}}) and (\protect{\ref{Rgauss}}) from
the MC simulations (`MC') and PCA simulations with
Poisson (`PCAP') and full (`PCAF') initial conditions.
X-axis rescaled and Y-axis
shifted for clarity.
For curve `$R_{MC}$', see text.
}
\label{gauss}
\end{figure}

This systematic error has no effect on the scaling behaviour, however.
If $F(x,t)=t^{b}\phi(x/t^a)$, we have
\begin{eqnarray}
\tilde F(x,t)&=&\frac{1}{2t\delta}\int_{t(1-\delta)}^{t(1+\delta)}(t')^b\phi
\left(\frac{x}{(t')^a}\right)dt'\\
&=&t^b\tilde \phi\left({x\over t^a}\right),
\end{eqnarray}
where $\tilde\phi(y)\equiv (2\delta)^{-1}\int_{1-\delta}^{1+\delta}\theta^b
\phi(y\theta^{-a})d\theta$, so $\tilde F$ has the same scaling properties
 as $F$.

In order to maximize the statistics, the reaction profile $R$ was measured
at every timestep between $t(1-\delta)$ and $t(1+\delta)$.  However, the
quantities $m$ and $l$ are much more cumbersome to measure using this
program (due to the multi-spin coding),
and so were only measured every 10 timesteps.  No significant loss in
statistics is incurred, since these quantities have very strong time
autocorrelations.  The {\sc Fortran} implementation of this algorithm
performed $3.7\times 10^{7}$ site updates per second on a HP 9000/715/75
workstation.

\subsection{Simulation Results}
\subsubsection{Poisson Initial Condition}
An initial condition with Poisson-like density fluctuations was prepared by
filling each of the appropriate site variables
(A particle for $x<0$, B particles for
$x>0$) with probability ${1\over 4}$.  The lattice size was 4000 sites, and
at the boundaries particles were free to leave the system,
with the density at
the boundary maintained at an average value of $1\over 2$ by
allowing A particles to enter from the left, and B particles to enter from the
right, randomly with probability $1\over 4$.  Measurements were taken at times
200-102400 timesteps, with the interval between measurements doubling
progressively.  The quantities $P_m(m)$, $P_m(l)$ and
$R(x,t)$ were measured as
described above, and then averaged over 82176 independent realizations of the
system.  The quantities $m^{(q)}$, $l^{(q)}$ and $x^{(q)}$ were then
measured from the $(1/q)$'th power of the normalized $q$'th moment of these
quantities.

\begin{figure}
\epsfxsize=\hsize
\centerline{\epsfbox{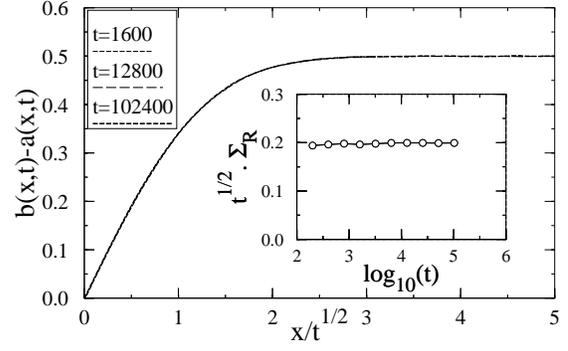}}
\caption{Scaling plot of the density difference
$a(x,t)$ $-b(x,t)$
for the
PCA data with random initial condition.  Inset: bias plot for the
total reaction rate $\Sigma_R\equiv\int R\,dx$.}\label{fse-pca}
\end{figure}

Figure \ref{fse-pca}
shows a plot of $(a-b)$ as a function of $(x/t^{1/2})$ for three time
values, and a plot of $\Sigma_R\cdot
t^{1/2}$ (where $\Sigma_R\equiv\int R\,dx$)
as a function of $t$.  These plots show that, just as in the
MC simulations, there are no finite size effects.

\begin{figure}
\epsfxsize=\hsize
\centerline{\epsfbox{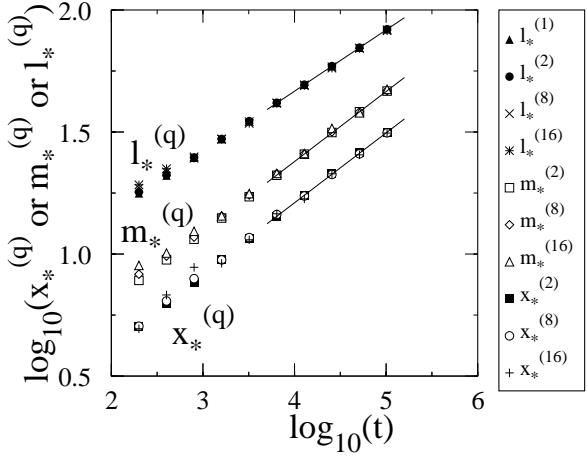}}
\caption{Log-log plot of $l_*^{(q)}$, $m_*^{(q)}$, and $x_*^{(q)}$
(see text) from the PCA simulations with Poisson initial conditions.
The curves for $m^{(q)}$ has been shifted vertically for clarity.}
\label{moms-pcar}
\end{figure}

Figure \ref{moms-pcar} is a log-log plot of $x_*^{(q)}$, $m_*^{(q)}$ and
$l_*^{(q)}$
as a function of time, where
\begin{equation}
x_*^{(q)}\equiv\mu_q x^{(q)},
\end{equation}
and $\mu_q$ is the appropriate scaling factor for Gaussian distributions
(see Eqs.\  (\ref{defmstar}) and (\ref{muq})).
The curves for $m_*^{(q)}$ have been shifted vertically (by 0.2) for clarity,
otherwise they would be too close to the curves for $x_*^{(q)}$.
The straight lines are a
fit to the last 5 points, for the lowest values of $q$.
The gradients of least-square fits for all the curves are summarized in Table
\ref{table}.  The collapse of the curves for different values of
$q$ confirms both the scaling hypothesis and the forms for the scaling
functions (\ref{pmgauss},\ref{plgauss}), and also that the reaction rate
profile has a Gaussian form:
\begin{equation}
R(x,t)={\Sigma_R\over w\sqrt{\pi}}\exp\left[-\left({x\over w}\right)^2\right]
\label{Rgauss}.
\end{equation}

\begin{figure}
\epsfxsize=\hsize
\centerline{\epsfbox{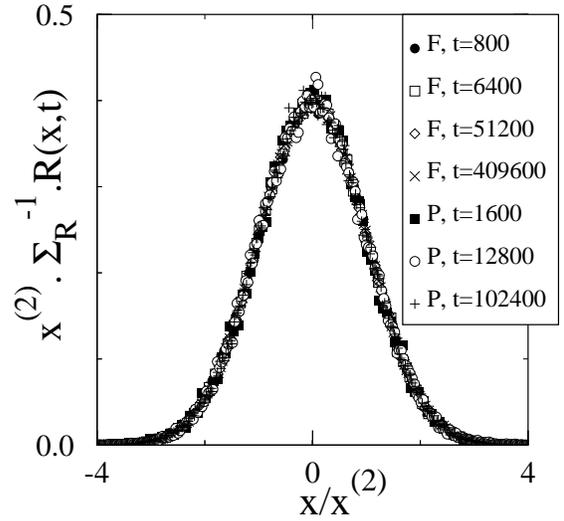}}
\caption{Scaling plot for $R(x,t)$, for the PCA simulations
with Poisson (`P') and full (`F') initial conditions.}
\label{pror}
\end{figure}

\begin{figure}
\epsfxsize=\hsize
\centerline{\epsfbox{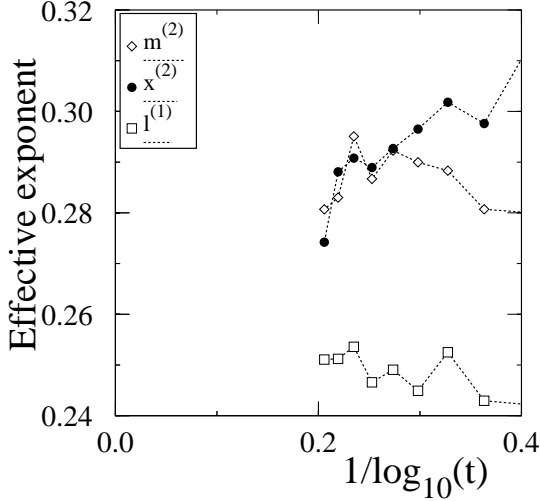}}
\caption{Effective exponents for $l^{(1)}$, $m^{(2)}$, and $x^{(2)}$
(see text) from the PCA simulations with Poisson initial conditions.}
\label{sg-pcar}
\end{figure}

Figure \ref{sg-pcar}
shows the effective exponents for
$x^{(2)}$, $m^{(2)}$ and $l^{(1)}$, from the successive gradients in
Fig.\ \ref{moms-pcar}.  The curves are much less noisy than those in
Fig.\ \ref{sg-mc}, by virtue of higher statistics and the use of coarse-grained
time averages.  There is a clear trend for the effective exponent
for $x^{(2)}$ to decrease as time increases, consistent with the
asymptotic value ${1\over 4}$ predicted elsewhere \cite{codr93,blee}.
The exponent for $m^{(2)}$ appears to increase initially, but the last few
points appear also to decrease, and in any case an asymptotic value $0.375$
is ruled out.

The rescaled forms of $P_l(l)$, $P_m(m)$, and $R$ are denoted by
`PCAP' in Figs.\
\ref{prol}, \ref{prom}, and \ref{pror} respectively.
Figure \ref{gauss} shows a fit of $P_l(l)$, $P_m(m)$ and $R$ to
the forms (\ref{pmgauss},\ref{plgauss},\ref{Rgauss}).
{}From Eq.\ (\ref{syserr}),
using $F(x,t)=At^{-\beta}\exp(-\lambda x^2/t^\alpha)$,
the fractional error introduced by the coarse-grained time averaging is
found to be $\approx (\delta^2/6)(x/w)^4$, where $w^2=(\int x^2 F\,dx/\int
F\,dx)$.  The measurement of these quantities is therefore
expected to be accurate for the first four decades or so, as is indeed
observed.

\subsubsection{Full Initial Conditions}
Because of the exclusion principle in the PCA model, the system is completely
static in regions where the occupation number is zero for one species and
assumes its maximal value for the other.
If one starts from a lattice that is filled with A-particles up to
$x=0$, and filled with B-particles for $x>0$,
simulations may be speeded up by only updating the lattice
in the region where a `hole' has penetrated.  By checking explicitly
that such holes never reach the physical boundary of the system, it is
possible to perform simulations on a system that is effectively infinite, so
having no finite-size effects.

\begin{figure}
\epsfxsize=\hsize
\centerline{\epsfbox{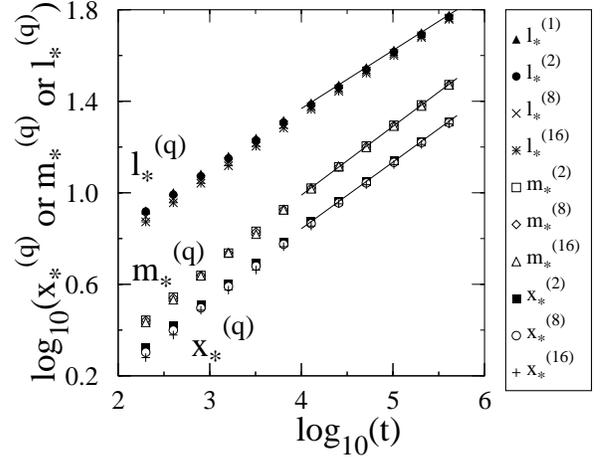}}
\caption{Log-log plot of $l_*^{(q)}$, $m_*^{(q)}$, and $x_*^{(q)}$
(see text) from the PCA simulations with full initial conditions.
The curves for $m^{(q)}$ has been shifted vertically for clarity.}
\label{moms-pcaf}
\end{figure}

\begin{figure}
\epsfxsize=\hsize
\centerline{\epsfbox{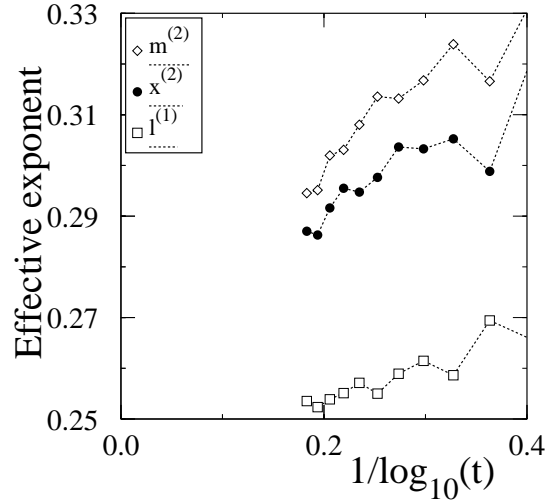}}
\caption{Effective exponents for  $l^{(1)}$, $m^{(2)}$, and $x^{(2)}$
(see text) from the PCA simulations with full initial conditions.}
\label{sg-pcaf}
\end{figure}

\begin{figure}
\epsfxsize=\hsize
\centerline{\epsfbox{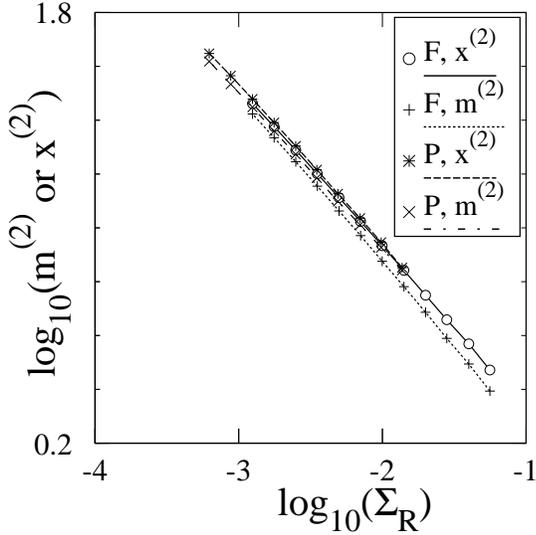}}
\caption{
Plot of $x^{(2)}$ and $m^{(2)}$ against the time-dependent current
for the two sets of PCA simulations.
}
\label{j-pca}
\end{figure}

Simulations of 64000 independent evolutions
of a full lattice were run for 409600 timesteps.
Measurements of $P_m(m)$, $P_l(l)$,
and $R$ were made using the
same method as for the Poisson initial condition.
Results for these simulations are shown in Figs. \ref{moms-pcaf}
and \ref{sg-pcaf}.
It might be expected
(considering the arguments in \cite{alhs}) that this case would be in a
different universality class from the case with
randomness in the initial state. However, the results for the
exponents (see Table \ref{table}) are very close to those measured for the
case of Poisson initial conditions,
and the marked decrease of the exponents for
$x^{(2)}$(arguably towards $0.25$)
is also seen in Fig.\ \ref{sg-pcaf}.
It is interesting to note that the transient trends in
$m^{(2)}$ and $l^{(1)}$ are in the opposite sense to the Poisson case.

Scaling plots for $P_l$, $P_m$ and $R$ are shown in Figs.\ \ref{prol},
\ref{prom}, and \ref{pror}, denoted by `PCAF'.
Plots of $l^{-1}P_l$, $P_m$, and $R$ may be found in Fig.\ \ref{gauss},
confirming that the profiles again
have the forms (\ref{pmgauss},\ref{plgauss},%
\ref{Rgauss}).

Figure \ref{j-pca} is a plot of $m^{(2)}$ and $x^{(2)}$
as a function of the time dependent current $\Sigma_R=\int R\,dx$, from the
simulations both with (`P') and without (`F')
Poisson fluctuations in the initial state.
The two curves for $x^{(2)}$ are almost coincident, which is what would be
expected
if the reaction profile depended upon the current only.  The curves for
$m^{(2)}$, however, are not quite coincident, showing that this quantity is
more sensitive to the initial condition.  Incidentally,
numerical tests showed that
the diffusion current at the origin has Poissonian noise
whether the initial state contained such
fluctuations or not.

\section{The Effect of Poissonian Fluctuations in the Initial State}
The measured value $m\sim t^{3/8}$ in Ref.\ \cite{alhs} was justified by
an argument about the Poisson fluctuations in the initial condition.
The argument went as follows: after time $t$, particles within a distance $\sim
t^{1/2}$ have had a chance of participating in the reaction.
The number of particles within a distance $t^{1/2}$ is of order
$t^{1/2}\pm c t^{1/4}$.
Since each reaction event kills precisely one A and one B,
there is therefore a local surplus $\sim t^{1/4}$ of one of the
species.  The majority species therefore invades the minority species by a
distance $m$, such that the number of minority particles between the origin and
$m$ is $\sim t^{1/4}$.  Since the particle profiles vary like
$x/t^{1/2}$, this means that $\int_0^m(x/t^{1/2})dx\sim
t^{1/4}$, so $m\sim t^{3/8}$

In order to assess the validity of this argument, it is possible to apply it
to a related quantity upon which analytical calculations may be made.
Consider the diffusion equation $\partial_t\rho(x,t)={1\over 4}
\partial_x^2\rho(x,t)$
in one spatial dimension, with an initial condition that consists of
a random series of negative Dirac delta peaks for $x<0$ and positive
Dirac delta peaks for $x>0$.  That is,
\begin{equation}
\rho(x,t=0)=\sum_{i=1}^\infty \delta(x-x_i) - \sum_{i=1}^\infty
\delta(x+y_i),\label{initcon}
\end{equation}
where $x_i>0$, $y_i>0$.
If the intervals between the $x_i$ and $y_i$ have a Poisson distribution,
one has
\begin{eqnarray}
\langle\rho(x,0)\rangle &=& \hbox{sign}(x),\\
\langle\rho(x,0)\rho(y,0)\rangle &=&\hbox{sign}(xy)+\delta(x-y),
\end{eqnarray}
where $\langle\rangle$ represents an average over the variables $x_i$, $y_i$.

The solution for $\rho$ may be written in the form
\begin{eqnarray}
\rho(x,t)&=&\int_{-\infty}^\infty \rho(x',0) (\pi t)^{-{1/2}}
\exp\left({-{(x-x')^2\over t}}\right)dx'\\
&=&\sum_i (\pi t)^{-{1/2}}\left[ \exp\left({-{(x-x_i)^2\over t}}\right)
\right. \nonumber\\
&&\phantom{\sum_i (\pi t)^{-{1/2}}\left[\right.}
\left.-\exp\left({-{(x+y_i)^2\over t}}\right)\right].\label{rhosum}
\end{eqnarray}

Consider the gradient of $\rho$:
\begin{eqnarray}
\partial_x\rho&=&\sum_i 2(\pi t)^{-{1/2}}
\left[ {x-x_i\over t}\exp\left({-{(x-x_i)^2\over t}}\right)\right.\nonumber\\
&&\phantom{ \sum_i 2(\pi t)^{-{1/2}}
\left[\right.}\left.
-{x+y_i\over t}\exp\left({-{(x+y_i)^2\over t}}\right)\right]\\
&=& 2{x\over t}\rho(x,t) \nonumber\\
&&+ \sum_i (\pi t^3)^{-{1/2}}\left[
x_i\exp\left({-{(x-x_i)^2\over t}}\right)\right.\nonumber\\
&&\phantom{+ \sum_i (\pi t^3)^{-{1/2}}\left[\right.}
\left.+y_i\exp\left({-{(x+y_i)^2\over t}}\right)\right].\label{zerograd}
\end{eqnarray}
The second term on the right hand side of Eq.\ (\ref{zerograd})
is strictly positive.  The gradient of $\rho$ when $\rho$ is zero
is therefore strictly positive, so, since $\rho$ is continuous for all
$t>0$, $\rho$ is zero at precisely one point, $x_0(t)$ (say).

It is possible to find the probability distribution of these zeros,
$P(x_0)$, over the ensemble of initial states.  The position $x_0$ is defined
by $\rho(x_0,t)=0$, or, equivalently,
\begin{equation}
\int_{-\infty}^\infty\rho(z,0)\exp\left(
{-{z^2\over t}}\right)\exp\left({-{2zx_0\over t}}
\right)dz=0.
\label{defx0}
\end{equation}
Suppose that $x_0\sim t^a$, where $a$ is expected to be
less that ${1\over 2}$, and let $\epsilon=t^{b+(1/2)}$, with
$0<b<({1\over 2}-a)$.
Then the contribution to the integral in (\ref{defx0})
for $|x|>\epsilon$ is of order
$\exp(-\epsilon^2/t)\sim \exp(-t^{2b})$, which vanishes as $t\to\infty$.
However, for $|x|<\epsilon,$
the argument of the second exponential has upper bound
$2\epsilon x_0/t\to 0$, and so the asymptotic value of the
integral is found by using the first few terms only
of the Taylor expansion of  this exponential.
In other words, the leading contribution to
$x_0$ as $t\to\infty$ is given by
\begin{equation}
\int_{-\infty}^\infty\rho(z,0)e^{-{z^2\over t}}dz
-{2x_0\over t}\int_{-\infty}^\infty\rho(z,0)z\,e^{-{z^2\over t}}dz
=0.
\end{equation}

The expectation value of the second moment of $x_0$ is
\begin{equation}
\langle x_0^2\rangle
=
{t^2\over 4} \left\langle
{\int \rho(x,0)e^{-{x^2\over t}}dx\over \int x\rho(x,0)e^{-{x^2\over t}}dx}
{\int \rho(y,0)e^{-{y^2\over t}}dy\over \int y\rho(y,0)e^{-{y^2\over t}}dy}
\right\rangle
\end{equation}
To evaluate this average, write $\rho(x,0)=\hbox{sign}(x) + \tau(x)$,
where $\langle\tau(x)\rangle=0$ and $\langle\tau(x)\tau(y)\rangle
=\delta(x-y)$.  Then $\int x\rho(x,0)\exp({-{x^2/ t}})dx = t
+\int x\tau(x)\exp({-{x^2/ t}})dx$, the second term being typically
much smaller than the first.  To find the leading contribution to
$x_0$, it is sufficient to replace $\int x\rho(x,0)\exp({-{x^2/ t}})dx$
in the
denominator by $t$.  We therefore have
\begin{eqnarray}
\langle x_0^2\rangle &=& {1\over 4}\int_{-\infty}^\infty\int_{-\infty}^\infty
\langle\rho(x,0)\rho(y,0)\rangle\nonumber\\
&&\phantom{{t\over 4}\int_{-\infty}^\infty\int_{-\infty}^\infty
\langle}\times    \exp\left({-{x^2+y^2\over t}}
\right)dxdy + \dots\\
&=& {1\over 4}\int_{-\infty}^\infty
e^{-{2x^2\over t}}dx + \dots\\
&=&{1\over 2}\sqrt{\pi t\over 8} + \dots\\
\end{eqnarray}
Similarly, the $2n$'th moment of $P(x_0)$ is of the form
\begin{eqnarray}
\langle x_0^{2n}\rangle &=&
{1\over 2^{2n}} \int_{-\infty}^\infty dx_1\dots\int_{-\infty}^\infty dx_{2n}
\nonumber\\
&&\phantom{{t\over 2}^n \int_{-\infty}^\infty }\times
\exp\left(-{\sum_{i=1}^{2n}x_i^2\over t}\right)\nonumber\\
&&\phantom{{t\over 2}^n \int_{-\infty}^\infty \times}\times
\langle\rho(x_1,0)\dots\rho(x_{2n},0)\rangle + \dots\\
&=& { (2n)!\over 2^{2n}n!}\left({\pi t\over 8}\right)^{n/2}+
\dots\label{x0moms}
\end{eqnarray}
For a distribution of the form $P(x_0)=\sqrt{(\lambda/ \pi)}\exp(-\lambda
x_0^2)$, one has
\begin{equation}
\langle x_0^{2n}\rangle = {(2n)!\over 2^{2n}\lambda^n n!}.
\end{equation}
Comparison with (\ref{x0moms}) gives
\begin{equation}
P(x_0)={2\over\pi}\sqrt{2\over t}\exp\left(-\sqrt{8\over\pi t}x_0^2\right).
\end{equation}
The distribution of $x_0$ is therefore characterized by a single lengthscale
$\lambda^{-{1/2}}\propto t^{1/4}$.

\begin{figure}
\epsfxsize=\hsize
\centerline{\epsfbox{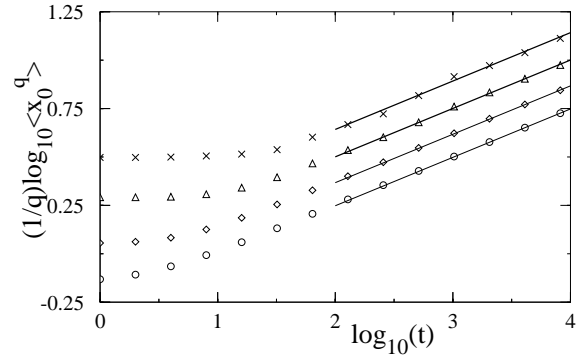}}
\caption{
Log-log plot of $\langle x_0^{q}\rangle$ versus $t$ for
$q=2$ ($\circ$), 4 ($\diamond$), 8 ($\triangle$) and 16 ($\times$), from
numerical solutions of
Eq.\ (\protect{\ref{rhosum}}).
The straight lines are the
asymptotic solutions from Eq.\  (\protect{\ref{x0moms}}).
\label{diff}}
\end{figure}

Figure \ref{diff} shows
the moments of $P_0$, averaged over 10000 realizations, from a
numerical solution of the
zero of $\rho$ from Eq.\  (\ref{rhosum}),
compared with the asymptotic predictions
of Eq.\  (\ref{x0moms}).

{}From Eq.\ (\ref{eom}), the density difference $(a-b)$ in the
reaction-diffusion
problem, averaged over evolutions, obeys a simple diffusion equation.
The quantity $\rho$ with the initial condition (\ref{initcon}) is therefore
equal to $(a-b)$ for the initial condition with Poisson fluctuations used
in the numerical simulations, with negative peaks corresponding to A particles
and positive peaks corresponding to B particles.
  The quantity $x_0$ differs from $m(t)$ because
the latter contains further fluctuations due to the diffusive noise that has
been averaged over in the former.  However, the argument used in \cite{alhs}
to obtain $m\sim t^{3/8}$ may be applied equally well to $x_0$.
The reaction centre shifts to compensate for a local majority of order
$t^{1/4}$ in one of the species, and the argument predicts $x_0\sim
t^{3/ 8}$.
It is interesting to note that the
correct exponent is obtained if the
initial value $a(x,0)=a_0$ is used instead of the value $a(x,t)\propto
{x/ t^{1/2}}$ at time $t$ in the balance equation
$\int^{x_0}a(x,t)\sim t^{1/4}$.  This ambiguity is probably the
reason for the argument being incorrect.

\begin{table}
\caption{
Comparison of the simulation results in this paper
for the Monte-Carlo model (MC) and PCA model with Poisson (PCAP) and Full
(PCAF) initial conditions with those of Araujo {\it et al\/}
(ALHS){\protect\cite{alhs}}.
Numbers in parentheses represent the statistical error in the preceding
digit.
}
\label{table}
\begin{tabular}{lllll}
& ALHS & MC & PCAP & PCAF \\
[0.05 cm]\hline
\rule[-0.3cm]{0.0cm}{0.0cm}
Size & 2000 & 4000 &  4000 & Infinite \\
Exclusion prle.? & No & No\tablenote{See text} & Yes & Yes\\
Initial density  & 1.0 & 1.0 & 0.5 & 2.0 \\
Initial State & (Uniform) & Poisson & Poisson & Uniform \\
Averaging & 6000--15000 & 21000 & 82176 & 64000 \\
Max time & 25000& 25000 & 102400 & 409600\\
Exponents: & & && \\
\hskip 0.5 cm $l^{(1)} $&0.25 &0.251(3) &0.2510(6)& 0.2542(4) \\
\hskip 0.5 cm $l^{(16)}$& 0.25&0.23(1) &0.248(3)&0.2609(3) \\
\hskip 0.5 cm $m^{(2)} $& 0.375&0.281(4) &0.287(1) &0.300(1) \\
\hskip 0.5 cm $m^{(16)}$&0.375 &0.29(1) &0.284(1) &0.299(3) \\
\hskip 0.5 cm $x^{(2)} $&0.312 &0.2799(2)\tablenote{Measured from $X^{(q)}$}
& 0.286(2)&0.291(1) \\
\hskip 0.5 cm $x^{(8)} $&0.359 &0.282(2)$^{\rm b}$ &0.280(4)&0.293(1) \\
\hskip 0.5 cm $x^{(16)}$&0.367 &0.30(2)$^{\rm b}$ &0.28(1)& 0.293(2)\\
Fit over last\dots & N/A & 8 points  & 5 points  & 6 points \\[.2cm]
\end{tabular}
\end{table}

\section{Conclusions}
It appears from extensive simulations that the reaction profile in this
system has the same simple dynamic scaling form, independently
of the presence of an exclusion principle and of randomness in the
initial state.
The motion of the reaction centre due to the Poisson noise
appears only to account for a contribution of order $\sim t^{1/4}$
to the reaction width, which is not large enough to alter the scaling
behaviour.
The measured exponent $\approx 0.29$
describing both the reaction width
and the midpoint fluctuations
appears in fact to be decreasing slowly in time, with
favourable evidence for an asymptotic value $0.25$.  This,
together with the measured form for the reaction profile, is consistent with
the steady-state results being applicable\cite{codr93,blee,codr94}.

It is, nevertheless,
surprising that the approach to the asymptotic behaviour should be so slow.
It is not clear whether logarithmic corrections should be present,
as they do not occur in the steady-state problem\cite{blee}.
However, in these simulations the ratio of the reaction width
$w$ to the
diffusion length $(Dt)^{1/2}$ was never smaller than
$\approx 0.2$, whereas the application of the
steady-state argument requires that this ratio be small.  This could account
for the fact that the asymptotic regime has not been reached.  Simulations
where this ratio is truly small would not appear to be practical at present.

An investigation of the simulation procedure used by Araujo {\it et al\/}
has revealed a few errors in the results published in \cite{alhs}.
A repeat of their simulations
appears to confirm the results of the present article for $P_m$ and
$P_l$, and the behaviour $m\sim t^{0.30}$, but does
not find that $R$ satisfies a scaling ansatz\cite{privcom}.
This inconsistency between my results and those of Araujo {\it et al\/}
is currently unexplained.

A recent calculation by Rodriguez and Wio\cite{rowio}
suggests that the reaction profile $R$ should be the superposition of two
scaling forms, with width exponents ${1\over 3}$ and ${3\over 8}$ respectively.
However,
these exponents and the form they predict for $R$
($\sim\exp[-(x/w)^{3/2}]$) do not agree with the results of simulations.
The approximation scheme they used would therefore not appear
to be valid,
unless it describes a regime inaccessible to simulations.

The simulation evidence in favour of dynamic scaling in this
model is very strong.  However,
the numerical evidence that all lengthscales scale asymptotically as $t^{1/4}$
is far from conclusive, and so needs to be put on a sound
theoretical basis, either by an exact calculation or by a rigorous
justification for the analogy with the static case.
\section*{Acknowledgements}
I would like to thank Michel Droz, Hernan Larralde, John Cardy, and Ben Lee for
many interesting discussions, and Michel Droz for a careful reading of this
manuscript.
I would also like to thank Mariela Araujo for making
details of her simulations available to me.


\begin{references}
\bibitem[*]{curad} Address from January 1995:
Department of Physics,
University of Guelph, Guelph, Ontario N1G 2W1, Canada.
\bibitem  {gara}
L.\ G\'alfi and Z.\ R\'acz, Phys.\ Rev.\ A {\bf 38},
3151 (1988).
\bibitem  {koliko} Y.-E.\ Koo, L.\ Li, and R.\ Kopelman, Mol.\
Cryst.\ Liq. Cryst. {\bf 183}, 187 (1990).
\bibitem  {jieb} Z.\ Jiang and C.\ Ebner, Phys.\ Rev.\ A
{\bf 42}, 7483 (1990).
\bibitem{chdr} B.\ Chopard and M.\ Droz, Europhys.\ Lett.\ {\bf 15} 459 (1991).
\bibitem {codrch91} S.\ Cornell, M.\ Droz, B.\ Chopard,
Phys.\ Rev.\ A {\bf 44}, 4826 (1991).
\bibitem  {koko} Y.-E.\ Koo and R.\ Kopelman, J.\ Stat.\ Phys.\
{\bf 65}, 893 (1991).
\bibitem {tahakitrwe} H.\ Taitelbaum, S.\ Havlin, J.\ Kiefer,
B.\ Trus, and G.\ Weiss,  J.\ Stat.\ Phys {\bf 65}, 873 (1991).
\bibitem {arhalast} M.\ Araujo, S.\ Havlin, H.\ Larralde, H.E.\
Stanley,  Phys.\ Rev.\ Lett. {\bf 68} 1791 (1992).
\bibitem{bnre} E.\ Ben-Naim and S.\ Redner, J.\ Phys.\ A {\bf 25}, L575 (1992)
\bibitem  {laarhast1}H.\ Larralde, M.\ Araujo, S.\ Havlin,
and H.\ E.\
Stanley, Phys.\ Rev.\ A {\bf 46}, 855 (1992).
\bibitem  {laarhast2} H.\ Larralde, M.\ Araujo, S.\ Havlin,
and H.\ E.\
Stanley, Phys.\ Rev.\ A {\bf 46} 6121 (1992).
\bibitem  {codrch92} S.\ Cornell, M.\ Droz, B.\ Chopard,
Physica {\bf A188}, 322 (1992).
\bibitem {takohakowe} H.\ Taitelbaum, Y.-E.\ Koo, S.\ Havlin, R.\ Kopelman,
and G.\ Weiss, Phys\ Rev.\ A {\bf 46}, 2151 (1992).
\bibitem{chdrkara} B.\ Chopard, M.\ Droz, T.\ Karapiperis, and Z.\ Racz,
Phys.\ Rev.\ E {\bf 47}, 40 (1993).
\bibitem{codr93} S.\ Cornell and M.\ Droz, Phys.\ Rev.\ Lett.\ {\bf 70}, 3824
(1993).
\bibitem{alhs} M.\ Araujo, H.\ Larralde, S.\ Havlin, and H.E.\ Stanley,
Phys.\ Rev.\ Lett.\ {\bf 71} 3592 (1993).
\bibitem{blee} B.\ Lee and J.\ Cardy, Phys.\ Rev.\ E, to appear (1994).
\bibitem{co94} S.\ Cornell, Phys.\ Rev.\ Lett.\, to appear (1994).
\bibitem{chdr91} B.\ Chopard and M.\ Droz, J.\ Stat.\ Phys.\ {\bf 64}, 859
(1991).
\bibitem{codr94} S.\ Cornell and M.\ Droz, in preparation (1994).
\bibitem{privcom} M.\ Araujo, private communication.
\bibitem{rowio} Rodriguez and Wio, preprint, (1994).
\end{references}
\end{document}